\def\etal{{\it et al.}}

\def\~{{$\tilde{\phantom{a}}$}}

\documentclass [12pt] {article}
\usepackage{epsfig}
\usepackage{color}

\def\thebibliography#1{\section{References}\markboth
 {REFERENCES}{REFERENCES}\list
 {[\arabic{enumi}]}{\settowidth\labelwidth{[#1]}\leftmargin\labelwidth
 \advance\leftmargin\labelsep
 \usecounter{enumi}}
 \def\newblock{\hskip .11em plus .33em minus -.07em}
 \sloppy
 \sfcode`\.=1000\relax}
\def\upcite#1{\raise6pt\hbox{\scriptsize
\cite{#1}}}
  \def\lsim{\mathrel {\vcenter {\baselineskip 0pt \kern 0pt
    \hbox{$<$} \kern 0pt \hbox{$\sim$} }}}
    \def\gsim{\mathrel {\vcenter {\baselineskip 0pt \kern 0pt
    \hbox{$>$} \kern 0pt \hbox{$\sim$} }}}

\def\hline{\noalign{\hrule \vskip2pt}}

\def\|{\ifmmode\Vert\else \char`\|\fi}
\ifx\oldzeta\undefined                          % hasn't been done yet, so 
  \let\oldzeta=\zeta                            % save old definiton
  \def\zzeta{{\raise 2pt\hbox{$\oldzeta$}}}     % make new definition
  \let\zeta=\zzeta                              % and attatch it
\fi

\ifx\oldchi\undefined                           % hasn't been done yet, so 
  \let\oldchi=\chi                              % save old definiton
  \def\cchi{{\raise 2pt\hbox{$\oldchi$}}}       % make new definition
  \let\chi=\cchi                                % and attatch it
\fi

\def\frac#1#2{{#1 \over #2}}

\def\half{\ifinner {\scriptstyle {1 \over 2}}
   \else {1 \over 2} \fi}

\def\simge{\mathrel{%
   \rlap{\raise 0.511ex \hbox{$>$}}{\lower 0.511ex \hbox{$\sim$}}}}
\def\simle{\mathrel{
   \rlap{\raise 0.511ex \hbox{$<$}}{\lower 0.511ex \hbox{$\sim$}}}}

\def\buildchar#1#2#3{{\null\!                   % \null, cancel space
   \mathop#1\limits^{#2}_{#3}                   % #1, #2 above, #3 below
   \!\null}}                                    % cancel space, \null
\def\overcirc#1{\buildchar{#1}{\circ}{}}

\def\slashchar#1{\setbox0=\hbox{$#1$}           % set a box for #1 
   \dimen0=\wd0                                 % and get its size
   \setbox1=\hbox{/} \dimen1=\wd1               % get size of /
   \ifdim\dimen0>\dimen1                        % #1 is bigger
      \rlap{\hbox to \dimen0{\hfil/\hfil}}      % so center / in box
      #1                                        % and print #1
   \else                                        % / is bigger
      \rlap{\hbox to \dimen1{\hfil$#1$\hfil}}   % so center #1
      /                                         % and print /
   \fi}                                         %

\def\subrightarrow#1{%                          % #1 under arrow
  \setbox0=\hbox{%                              % set a box
    $\displaystyle\mathop{}%                    % no mathop
    \limits_{#1}$}%                             % just limits
  \dimen0=\wd0%                                 % get width
  \advance \dimen0 by .5em%                     % add a bit
  \mathrel{%                                    % space like =
    \mathop{\hbox to \dimen0{\rightarrowfill}}% % arrow to width
       \limits_{#1}}}                           % text below

\def\overlay#1#2{\ifmmode%
\setbox0=\hbox{$#1$}%
\setbox1=\hbox to\wd0{\hss$#2$\hss}\else%
\setbox0=\hbox{#1}%
\setbox1=\hbox to\wd0{\hss#2\hss}\fi%
#1\hskip-\wd0\box1 }

\def\pmb#1{\leavevmode\setbox0=\hbox{#1}%
\kern-.02em\copy0\kern-\wd0
\kern.04em\copy0\kern-\wd0
\kern-.02em\raise.04em\box0 }

\def\vereq#1#2{\lower3pt\vbox{\baselineskip1.5pt \lineskip1.5pt
\ialign{$\m@th#1\hfill##\hfil$\crcr#2\crcr\sim\crcr}}}

\def\tensor#1{\protect\@ontopof{#1}{\leftrightarrow}{1.15}\mathord{\box2}}
\def\overstar#1{\protect\@ontopof{#1}{\ast}{1.15}\mathord{\box2}}
\def\overdots#1{\protect\@ontopof{#1}{\cdots}{1.0}\mathord{\box2}}
\def\overcirc#1{\protect\@ontopof{#1}{\circ}{1.2}\mathord{\box2}}
\def\loarrow#1{\protect\@ontopof{#1}{\leftarrow}{1.15}\mathord{\box2}}
\def\roarrow#1{\protect\@ontopof{#1}{\rightarrow}{1.15}\mathord{\box2}}

\def\@ontopof#1#2#3{%
{\mathchoice
{\@@ontopof{#1}{#2}{#3}\displaystyle\scriptstyle}%
{\@@ontopof{#1}{#2}{#3}\textstyle\scriptstyle}%
{\@@ontopof{#1}{#2}{#3}\scriptstyle\scriptscriptstyle}%
{\@@ontopof{#1}{#2}{#3}\scriptscriptstyle\scriptscriptstyle}%
}%
}

\def\@@ontopof#1#2#3#4#5{%
\setbox0=\hbox{$#4#1$}%
\setbox1=\hbox{$#5#2$}%
\setbox2=\hbox{}\ht2=\ht0 \dp2=\dp0 %
\ifdim\wd0>\wd1 %
\setbox1=\hbox to\wd0{\hss\box1\hss}%
\mathord{\rlap{\raise#3\ht0\box1}\box0}%
\else   %
\setbox1=\hbox to.9\wd1{\hss\box1\hss}%
\setbox0=\hbox to\wd1{\hss$#4\relax#1$\hss}%
\mathord{\rlap{\copy0}\raise#3\ht0\box1}%
\fi
}%

\def\lambdabar{\protect\@lambdabar}
\def\@lambdabar{%
\relax
\bgroup
\def\@tempa{\hbox{\raise.73\ht0
\hbox to0pt{\kern.25\wd0\vrule width.5\wd0
height.1pt depth.1pt\hss}\box0}}%
\mathchoice{\setbox0\hbox{$\displaystyle\lambda$}\@tempa}%
{\setbox0\hbox{$\textstyle\lambda$}\@tempa}%
{\setbox0\hbox{$\scriptstyle\lambda$}\@tempa}%
{\setbox0\hbox{$\scriptscriptstyle\lambda$}\@tempa}%
\egroup
}

\def\corresponds{{\lower.2ex\hbox{=}}{\rm\kern-.75em^\triangle}}
\def\succsim{\succ\kern-.9em_\sim\kern.3em}
\def\precsim{\prec\kern-1em_\sim\kern.3em}
\def\slantfrac#1#2{\kern1em^{#1}\kern-.3em/\kern-.1em_{#2}}

\begin{document}

\begin{center}
{\Large\bf Slow light}
\\

\medskip

Kirk T.~McDonald
\\
{\sl Joseph Henry Laboratories, Princeton University, Princeton, NJ 08544}
\\
(April 3, 1999)
\end{center}

\section{Problem}

Consider a classical model of matter in which spectral lines are
associated with oscillators.  In particular, consider a gas with two
closely spaced spectral lines, $\omega_{1,2} = \omega_0 \pm \Delta/2$, where
$\Delta \ll \omega_0$.  Each line has
the same damping constant (and spectral width) $\gamma$, where
$\gamma \ll \Delta$.

\begin{figure}[htp]  % h = here, t = top, b = bottom, p = new page
\begin{center}
\includegraphics[width=4in]{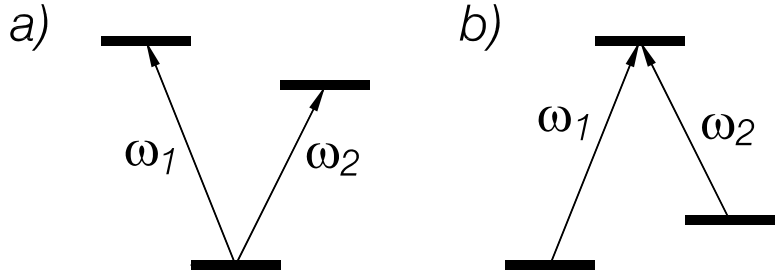}
\parbox{5.5in} % change 5.5in to \hsize for full-width caption
{\caption[ Short caption for table of contents ]
{\label{v-l} Two possible level diagrams for a three-level atomic system:
a) the ``V'' configuration; b) the ``$\Lambda$'' configuration.
}}
\end{center}
\end{figure}

Classically, one might expect these oscillators to correspond to the ``V''
configuration of atomic levels sketched in Fig.~\ref{v-l}a), in which both
higher levels can decay to the ground state by emission of photons of
frequencies
$\omega_1$ and $\omega_2$.  However, quantum mechanics also permits the
``$\Lambda$'' configuration sketched in Fig.~\ref{v-l}b) in which the highest
level can decay to the ground state via emission of a photon of frequency
$\omega_1$ as well as to the intermediate level via emission at frequency
$\omega_2$.  

The present problem is based on a ``$\Lambda$'' configuration in which a laser
of frequency $\omega_2$ pumps oscillator 2.  But since the lower level of
this oscillator is not the ground state, the pumping does not result in an
inverted population.  The physics of this system can be fairly well described
by a classical model in which the damping constant of the second oscillator
is taken to be negative: $\gamma_2 = - \gamma$.  The strengths of both
oscillators are positive.

Deduce an expression for the group velocity of a pulse of light centered
on frequency $\omega_0$ in this medium.  Show also that frequencies very near
$\omega_0$ propagate without attenuation.

In a recent experiment \cite{Hau}, the group velocity of light was reduced to
38 mph (17 m/s) by this technique
in a sodium vapor of density $N = 8 \times 10^{13}$ atoms/cm$^3$ using a pair 
of lines for which $\Delta / 2 \pi \approx 2 \times 10^7$/s.

\section{Solution}

 In a medium of index of refraction $n(\omega)$, the 
dispersion relation can be written
\begin{equation}
k = {\omega n \over c},
\label{eq1}
\end{equation}
where $k$ is the wave number and $c$ is the speed of light.
The group velocity is then given by
\begin{equation}
v_g = {d\omega \over dk} = {1 \over dk/d\omega} 
= {c \over n + \omega {dn \over d\omega}}.
\label{eq2}
\end{equation}

We next recall the classical oscillator model for the index of
refraction.  The index $n$ is the square root of the dielectric constant
$\epsilon$, which is in turn related to the atomic polarizability $\alpha$
according to (in Gaussian units)
\begin{equation}
D = \epsilon E = E + 4 \pi P = E(1 + 4 \pi N \alpha),
\label{eq3}
\end{equation}
where $D$ is the the electric displacement, $E$ is the electric field, $P$ is the polarization
density, and $N$ is the number density of atoms.  Then,
\begin{equation}
n = \sqrt{\epsilon} \approx 1 + 2 \pi N \alpha,
\label{eq4}
\end{equation}
for a dilute gas with index near 1.

The polarizability $\alpha$ is obtained from the dipole moment 
$p = ex = \alpha E$ induced by electric field $E$.  In the case of a single 
spectral line of frequency $\omega_0$, we say that the electron of
charge $e$ and mass $m$ is bound to the (fixed) nucleus by a spring of constant
$K = m \omega_0^2$, and the motion is subject to damping $-m \gamma \dot x$
where the dot indicates differentiation with respect to time.
The equation of motion in the presence of a wave of frequency $\omega$ is
\begin{equation}
\ddot x + \gamma \dot x + \omega_0^2 x = {e E \over m} = {e E_0 \over m}
 e^{i\omega t}.
\label{eq5}
\end{equation}
Hence,
\begin{equation}
x = {e E \over m} {1 \over \omega_0^2 - \omega^2 - i \gamma \omega}
= {e E \over m} {\omega_0^2 - \omega^2 + i \gamma \omega \over
(\omega_0^2 - \omega^2)^2 + \gamma^2 \omega^2},
\label{eq6}
\end{equation}
and so the polarizability is
\begin{equation}
\alpha = {e^2\over m} {\omega_0^2 - \omega^2 + i \gamma \omega \over
(\omega_0^2 - \omega^2)^2 + \gamma^2 \omega^2}.
\label{eq7}
\end{equation}

\begin{figure}[htp]  % h = here, t = top, b = bottom, p = new page
\begin{center}
\includegraphics[width=4in]{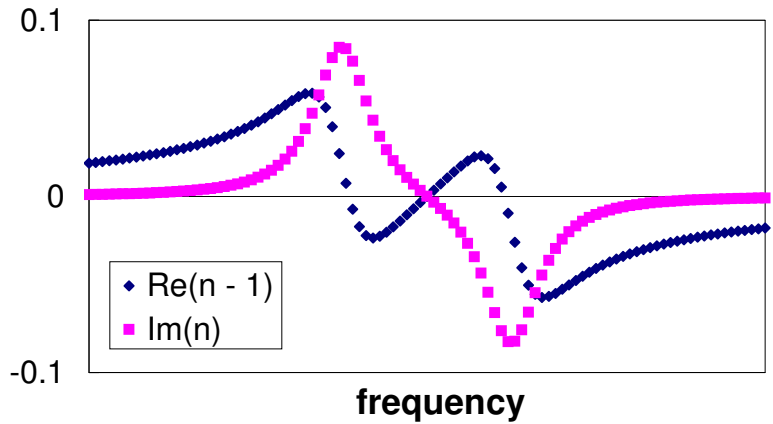}
\parbox{5.5in} % change 5.5in to \hsize for full-width caption
{\caption[ Short caption for table of contents ]
{\label{fig1} The real and imaginary parts of the index of refraction 
corresponding to the polarizability (\ref{eq4}) in a 
medium with one of a pair of spectral lines pumped so as to produce a large 
rate of change of the index with nearby frequency.
}}
\end{center}
\end{figure}

In the present problem, we have two spectral lines, $\omega_{1,2} = \omega_0
\pm \Delta/2$, both of unit oscillator strength, but line 2 is
pumped so that $\gamma_2 = - \gamma_1 = -\gamma$.  In this case, the
polarizability is given by
\begin{eqnarray}
\alpha & = & {e^2\over m} {(\omega_0 - \Delta/2)^2 - \omega^2 
+ i \gamma \omega \over
((\omega_0 - \Delta/2)^2 - \omega^2)^2 + \gamma^2 \omega^2}
+ {e^2\over m} {(\omega_0 + \Delta/2)^2 - \omega^2 
- i \gamma \omega \over
((\omega_0 + \Delta/2)^2 - \omega^2)^2 + \gamma^2 \omega^2}
\nonumber \\
& \approx &  {e^2\over m} {\omega_0^2 - \Delta \omega_0 - 
\omega^2 + i \gamma \omega \over
(\omega_0^2 - \Delta \omega_0 - \omega^2)^2 + \gamma^2 \omega^2}
+ {e^2\over m} {\omega_0^2 + \Delta \omega_0 - \omega^2 
- i \gamma \omega \over
(\omega_0^2 + \Delta \omega_0 - \omega^2)^2 + \gamma^2 \omega^2},
\label{eq8}
\end{eqnarray}
where the approximation is obtained by the neglect of 
terms in $\Delta^2$ compared to those in $\Delta \omega_0$.
The index of refraction (\ref{eq4}) corresponding to polarizability 
(\ref{eq8}) is shown in Fig.~\ref{fig1}.

We next consider the issue of attenuation of a pulse of frequency $\omega$.
Since $k = \omega n/c \approx \omega (1 + 2 \pi N \alpha)/c$, the spatial 
dependence $e^{ikz}$ of a pulse propagating in the $z$ direction includes 
attenuation
if the imaginary part of the index $n$ is positive.  However, the pumping
described by $\gamma_2 = -\gamma_1$ leads to $Im[\alpha(\omega_0)]
= 0$.  Hence, there is no attenuation of a probe pulse at frequency $\omega_0$.
This phenomenon has been called electromagnetically induced transparency
\cite{Harris}.  

Since $\alpha(\omega_0) = 0$, we have $n(\omega_0)
= 1$, and the phase velocity at $\omega_0$ is exactly $c$.

To obtain the group velocity at frequency $\omega_0$, we need the derivative
\begin{equation}
{dRe(n) \over d\omega} \biggr|_{\omega_0} =
2 \pi N {dRe(\alpha) \over d\omega} \biggr|_{\omega_0}
= {4 \pi N e^2 (\Delta^2 - \gamma^2) \over m \omega_0 (\Delta^2 + \gamma^2)^2}
\approx {4 \pi N e^2 \over m \omega_0 \Delta^2},
\label{eq9}
\end{equation}
where the approximation holds when $\gamma \ll \Delta$ as is the case here.
  For large density $N$, the group velocity (\ref{eq2}) is given by
\begin{equation}
v_g \approx {\Delta^2 \over 4 \pi N r_0 c},
\label{eq10}
\end{equation}
where $r_0 = e^2/mc^2 \approx 3 \times 10^{-13}$ cm is the classical electron 
radius.  The group velocity is lower in a denser medium.  

In the experiment of Hau
\etal, the medium was sodium vapor, cooled to less than 100 nK to
increase the density.  An additional increase in density by a factor of 16
was obtained when the vapor formed a Bose condensate.  Plugging in the 
experimental parameters, $N  = 8 \times 10^{13}$/cm$^3$ and $\Delta \approx
1.2 \times 10^8$/s, we find $v_g \approx 1700$ cm/s as observed in the lab.

\end{document}